\documentclass[12pt]{article}
\usepackage{epsfig,cite,amsmath,amssymb,hhline,color}

\input{paperdef}
\graphicspath{{figs/}}

\evensidemargin \oddsidemargin
\newcommand{\mr}{\mathrm}
\newcommand{\bb}{$b\bar{b}$}

\newcommand{\sixoo}{60 \ifb}
\newcommand{\sixooeff}{60 \ifb\,eff$\times2$}
\newcommand{\sixooo}{600 \ifb}
\newcommand{\sixoooeff}{600 \ifb\,eff$\times2$}

\makeatletter
\renewcommand{\section}{\@startsection {section}{1}{\z@}%
                           {-3.5ex \@plus -1ex \@minus -.2ex}%
                           {2.3ex \@plus.2ex}%
                           {\mathversion{bold}\normalfont\Large\bfseries}}
\renewcommand{\subsection}{\@startsection{subsection}{2}{\z@}%
                           {-3.25ex\@plus -1ex \@minus -.2ex}%
                           {1.5ex \@plus .2ex}%
                           {\mathversion{bold}\normalfont\large\bfseries}}
\renewcommand{\subsubsection}{\@startsection{subsubsection}{3}{\z@}%
                           {-3.25ex\@plus -1ex \@minus -.2ex}%
                           {1.5ex \@plus .2ex}%
                           {\mathversion{bold}\normalfont\normalsize\bfseries}}
\makeatother

\begin{document}

\thispagestyle{empty}
\setcounter{page}{0}
\def\thefootnote{\fnsymbol{footnote}}

\begin{flushright}
\mbox{}
DCPT/11/54\\
DESY 11-101 \\
IPPP/11/27 \\
\end{flushright}

\vspace{1cm}

\begin{center}

{\large\sc {\bf Exclusive production of the BSM Higgs bosons at the LHC}}
\footnote{talk given by M.T.\ at the {\em DIS 2011}, 
April 2011, Newport News, Virginia, USA}

\vspace{1cm}

{\sc 
S.~Heinemeyer$^{1}$%
\footnote{
email: Sven.Heinemeyer@cern.ch}%
, V.A.~Khoze$^{2,3}$%
\footnote{
email: V.A.Khoze@durham.ac.uk}%
, M.G.~Ryskin$^{2,4}$%
\footnote{
email: misha.ryskin@durham.ac.uk}%
,\\[.5em] M.~Tasevsky$^{5}$%
\footnote{
email: Marek.Tasevsky@cern.ch}%
~and G.~Weiglein$^{6}$%
\footnote{email: Georg.Weiglein@desy.de}
}

\vspace*{1cm}

{\it
$^1$Instituto de F\'isica de Cantabria (CSIC-UC), 
Santander,  Spain\\

\vspace{0.3cm}
$^2$IPPP, Department of Physics, Durham University, 
Durham DH1 3LE, U.K.\\

\vspace{0.3cm}
$^3$School of Physics \& Astronomy, University of Manchester, 
Manchester M13 9PL, U.K.\\

\vspace{0.3cm}
$^4$Petersburg Nuclear Physics Institute, Gatchina, 
St.~Petersburg, 188300, Russia\\

\vspace{0.3cm}
$^5$Institute of Physics, ASCR, 
Na Slovance2, 18221 Prague, Czech Republic

\vspace{0.3cm}

$^6$DESY, Notkestra\ss e 85, D--22607 Hamburg, Germany
}
\end{center}

\vspace*{0.2cm}

\BC {\bf Abstract} \EC
We review the prospects for Central Exclusive Production (CEP) of BSM Higgs 
bosons at the LHC using forward proton detectors proposed to be installed at
220~m and 420~m from the ATLAS and/ or CMS. Results are presented for MSSM
in standard benchmark scenarios, in scenarios compatible with the
Cold Dark Matter relic abundance and other precision measurements, and
for SM with a fourth generation of fermions. We show that CEP can give a
valuable information about spin-parity properties of the Higgs bosons. 

\def\thefootnote{\arabic{footnote}}
\setcounter{footnote}{0}

\newpage




\section{Introduction}\label{Intro}
The central exclusive production (CEP)
of new particles has received a great deal of attention in recent years (see
\cite{FP420TDR} and references therein). 
The process is defined as $pp\rightarrow p\oplus\phi\oplus p$
and all of the energy lost by the protons during the interaction
(a few per cent) goes into the production of the central system, $\phi$. The 
final state therefore consists of a centrally produced system (e.g. dijet, 
heavy particle or Higgs boson) coming from a hard subprocess, two very forward 
protons and no other activity. The '$\oplus$' sign denotes the regions devoid 
of activity, often called rapidity gaps. Studies of the Higgs boson produced in
CEP form a core of the physics motivation for upgrade projects to install 
forward proton detectors at 220~m and 420~m from the ATLAS \cite{ATLASTDR} and
CMS \cite{CMSTDR} detectors, see \cite{FP420TDR} and \cite{AFPLOI}. Proving, 
however, that the detected central system is the Higgs boson coming from the 
SM, MSSM or other BSM theories will require measuring 
precisely its spin, \cp\ properties, mass, width and couplings.  

\section{Updates to the previous analyses}
\begin{figure}[h]
\includegraphics[width=.5\textwidth,height=5.5cm]{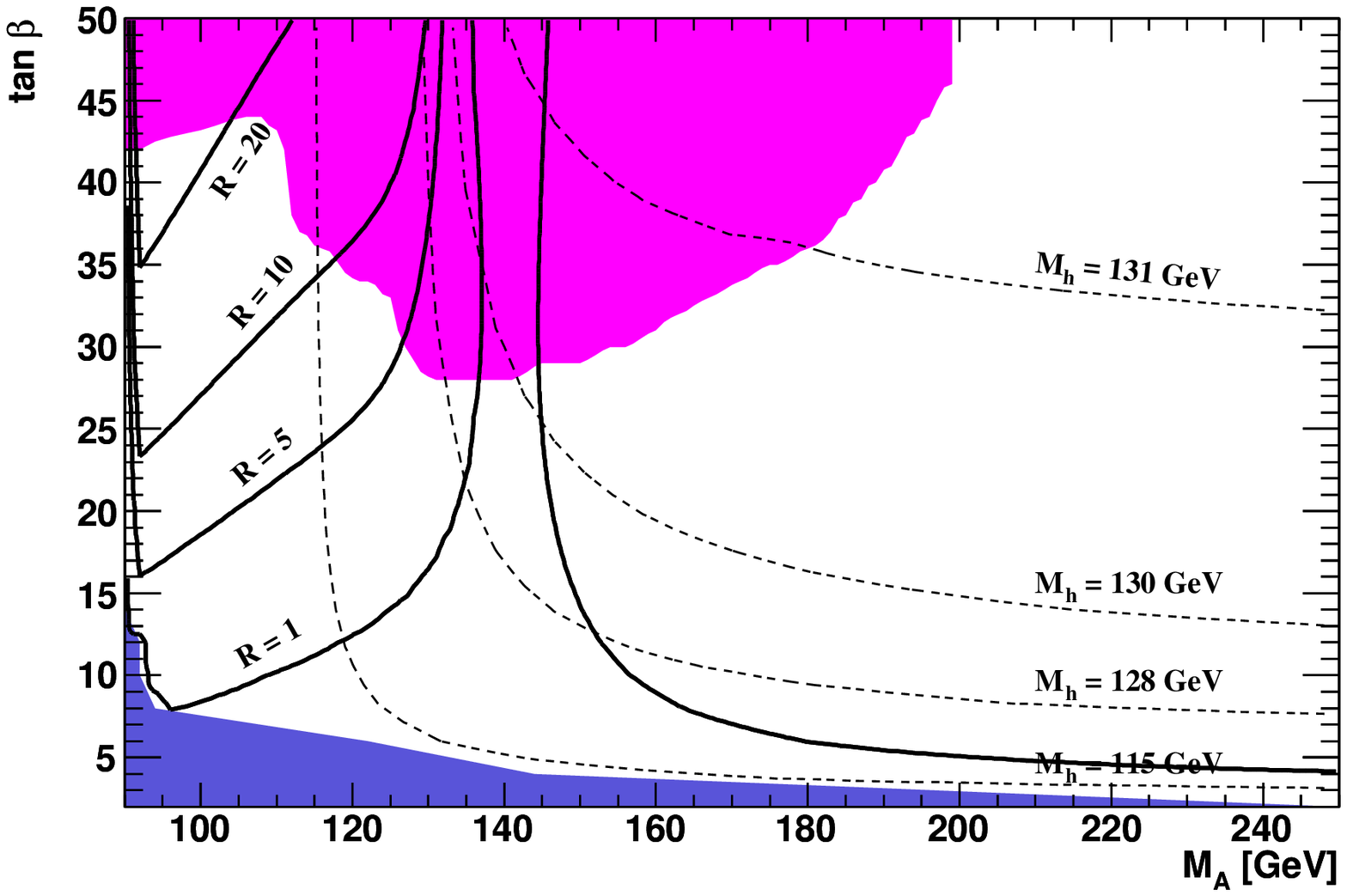}
\includegraphics[width=.5\textwidth,height=5.5cm]{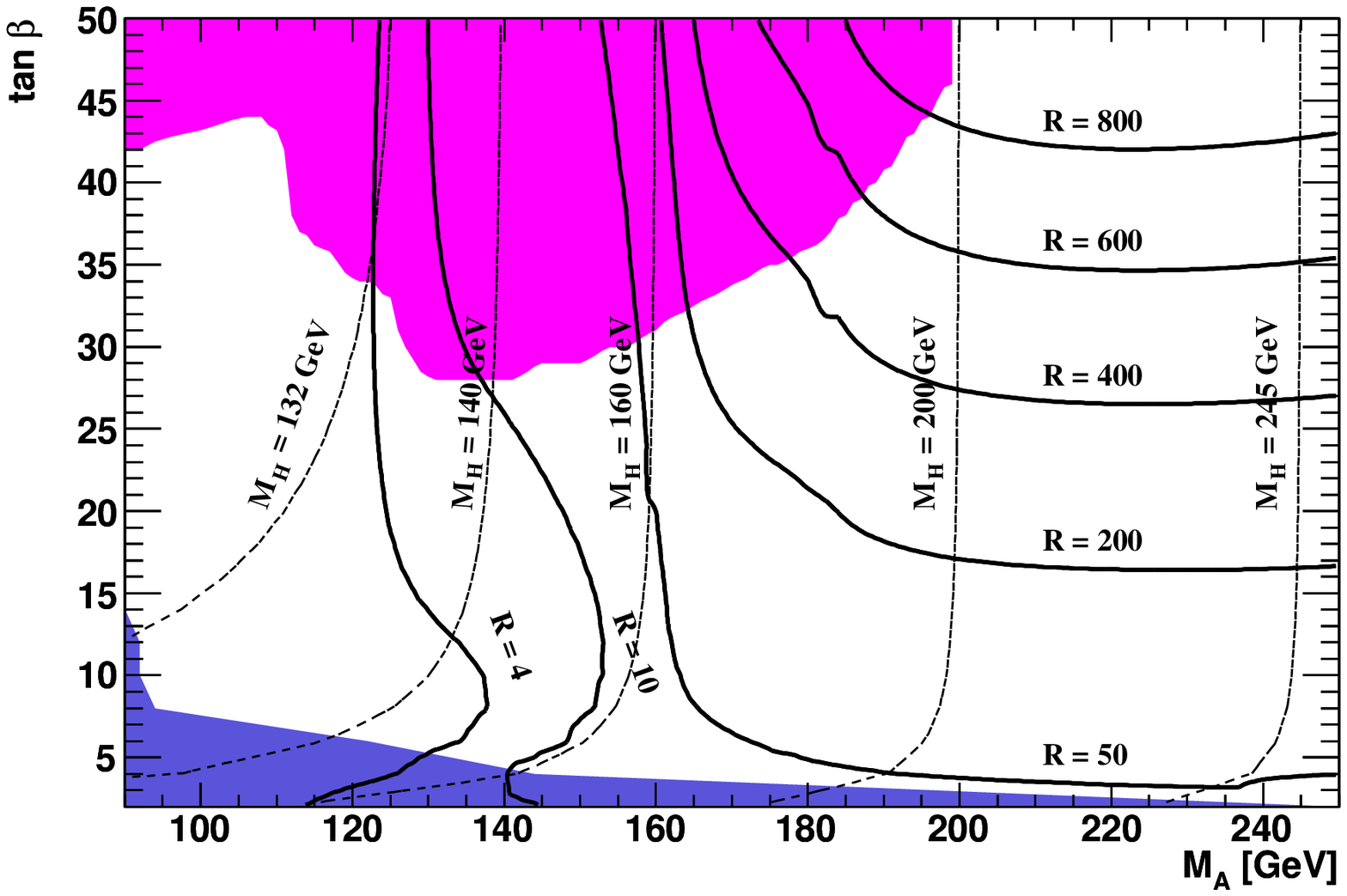}
\caption{Contours for the ratio of signal events in the MSSM to those in the SM
and for the mass values $M_h$ ($M_H$) for $h(H)\rightarrow b\bar{b}$ channel in
CEP are shown on left (right) for the $M_h^{\mr m\mr a\mr x}$ scenario with $\mu=200$~GeV. The 
dark (lighter) shaded region corresponds to the parameter region excluded by 
the LEP (Tevatron) Higgs boson searches.}  
\label{ratios}
\end{figure}

In \cite{HKRSTW} we have presented detailed results on signal and background 
predictions of CEP production (based on calculations in \cite{KMR}) of the 
light ($h$) and heavy ($H$) Higgs bosons. A recent update of results from 
\cite{HKRSTW} has been presented in \cite{HKRTW}. Changes between these two 
publications can be briefly summarized as:
\begin{itemize}
\item The NLO corrections added to the background associated with bottom-mass
terms in the Born amplitude \cite{Shuvaev} result in a suppression of the LO 
contribution by a factor of two or more for larger masses. 
\item The use of the recent version of FeynHiggs code \cite{feynhiggs}: all 
three main changes increase the bottom loop contribution 
and hence the $gg\rightarrow h(H)$ production rate:
the running of the bottom mass $m_b(m_b)$ rather than $m_b(m_t)$ in the bottom 
Yukawa coupling; the improved corrections to the bottom loop in the 
$\phi\rightarrow gg$ calculation; change to a running top mass, efectively 
decreasing the top-loop contribution.  
\end{itemize}
These changes result in enlarging the regions covered by 5$\sigma$ or 3$\sigma$
contours compared to those in \cite{HKRSTW}. The change in the 
signal cross section is visualised as ratios of the MSSM to SM cross sections 
shown in Fig.~\ref{ratios}, and has to be compared with Figs. 2 and 7 of 
\cite{HKRSTW}. We conclude that the MSSM cross section increased 
at lower $M_A$ (all $M_A$) for $h$ $(H)$.

Four luminosity scenarios are considered: ``\sixoo'' and
``\sixooo'' refer to running at low and high instantaneous luminosity, 
respectively, using conservative assumptions for the signal rates and the
experimental efficiencies (taken from \cite{CMS-Totem}); possible improvements
on the side of theory and experiment could allow for scenarios where the event
rates are enhanced by a factor 2, denoted by ``\sixooeff'' and ``\sixoooeff''.
\section{Cold Dark Matter benchmark scenarios}
\begin{figure}[h]
\includegraphics[width=.5\textwidth,height=5.5cm]{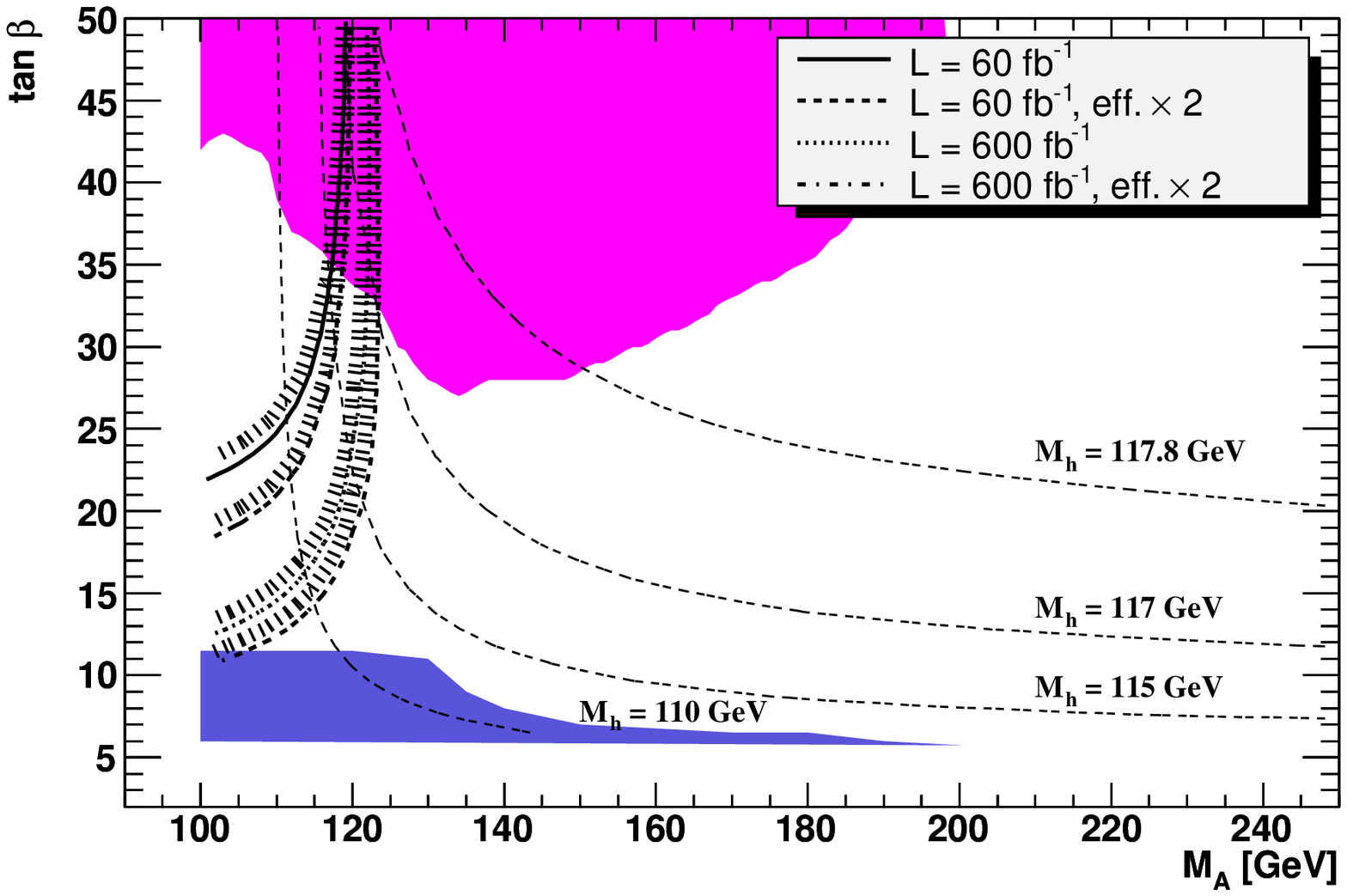}
\includegraphics[width=.5\textwidth,height=5.5cm]{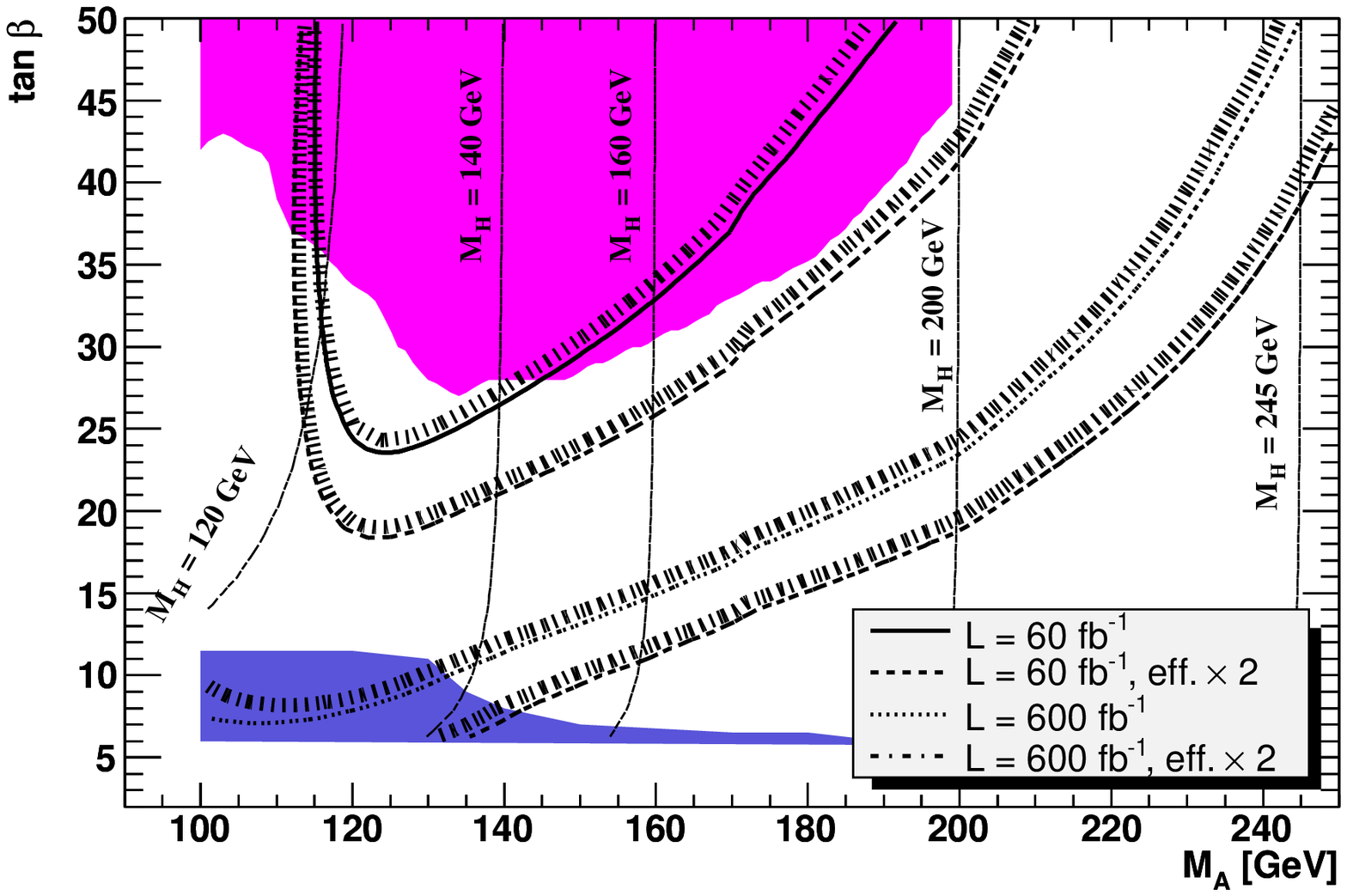}
\caption{5$\sigma$ discovery and mass $M_h$ ($M_H$) contours for 
$h(H)\rightarrow$ \bb\ channel in CEP production in the 
$M_A-\tan\beta$ plane of the MSSM are shown on left (right) within the CDM 
benchmark scenario {\bf P3}. The results are shown for four assumed effective 
luminosities (see the text). The dark (lighter) shaded region corresponds to 
the parameter region excluded by the LEP (Tevatron) Higgs boson searches.}  
\label{CDM}
\end{figure}

Standard benchmark scenarios designed to highlight specific characteristics of 
the MSSM Higgs sector, so called $M_h^{\mr m\mr a\mr x}$ and no-mixing scenarios, do not 
necessarily comply with other than MSSM Higgs sector constraints. Scenarios 
which fulfill constraints also from electroweak precision 
data, B physics data and abundance of Cold Dark Matter (CDM) are the so called 
CDM benchmark scenarios \cite{CDM}. As observed and discussed in \cite{HKRTW}, 
the $5\sigma$
discovery and $3\sigma$ contours show in general similar qualitative features
as the results in the $M_h^{\mr m\mr a\mr x}$ and no-mixing scenario. In 
Fig.~\ref{CDM} the $5\sigma$ discovery contours are shown for the \bb\ decay 
channel in the P3 plane. For the ligh Higgs boson $h$, a $5\sigma$ discovery 
is possible for $M_A \lesssim$ 125~GeV and $\tan\beta\gtrsim$10, depending on 
luminosity. The LEP exclusion regions are observed to be complementary to the 
parameter space covered by CEP Higgs boson production. For the heavy Higgs 
boson $H$, the $5\sigma$ discovery can be reached up to $M_H \lesssim 260$~GeV 
at large $\tan\beta$ and high luminosity. At low luminosity, the reach extends 
only up to $M_H \lesssim 210$~GeV, and it is largely excluded by the Tevatron 
searches.
\section{Model with a fourth generation of fermions}
A rather simple example of physics beyond SM is a model ``SM4'' which extends 
the SM by a fourth generation of heavy fermions, see for instance \cite{SM4}. 
The masses of the fourth generation quarks in such a scenario need to be 
significantly larger than the mass of the top quark. As a consequence, the 
effective coupling of the Higgs boson to two gluons in the SM4 is to good 
approximation three times larger than in the SM and the partial decay width 
$\Gamma(H\to gg)$ are larger by a factor of 9, giving rise to a corresponding 
shift in the total Higgs width and therefore all the decay branching ratios, 
see for instance \cite{four-gen-and-Higgs}. The total decay width in the SM4 
and the relevant decay branching ratios in terms of the corresponding 
quantities in the SM have been evaluated in \cite{HKRTW}. 
Recent combined analyses of the CDF and D\O\ collaborations \cite{SM4-CDF-D0},
and the LEP Higgs searches \cite{LEPHiggs} (data from \cite{LEPHiggs} were 
re-interpreted using the HiggsBounds program \cite{higgsbounds}) exclude Higgs 
bosons of the 
SM4 at the 95\% C.L. in regions 130~GeV $\lesssim M_{H^{SM4}} \lesssim$ 210~GeV
and $M_{H^{SM4}} \lesssim 112$~GeV, respectively.   

As discussed in \cite{HKRTW}, the \bb\ channel shows that even at
rather low luminosity the allowed region of 112~GeV $\lesssim M_{H^{SM4}} 
\lesssim$ 130~GeV can be covered by the CEP Higgs boson production. The still 
allowed region of $M_{H^{SM4}}>210$~GeV cannot be covered due to a 
low BR($H^{SM4}\rightarrow$ \bb). The $\tau^+\tau^-$ channel in the allowed
mass region reaches a 
sensitivity of about $2\sigma$ at luminosity of \sixoo, while it can exceed 
$5\sigma$ at \sixooo.

\section{Coupling structure and spin-parity determination}
Standard methods to determine the spin and the \cp\ properties of Higgs bosons 
at the LHC rely to a large extent on the coupling of a relatively heavy Higgs 
boson to two gauge bosons. In particular, the channel $H\rightarrow 
ZZ\rightarrow\!4l$ - if it is open - offers good prospects in this respect
\cite{HZZ}. In a 
study \cite{Ruwiedel} of the Higgs production in the weak vector boson 
fusion it was found that for $M_H=160$~GeV the $W^+W^-$ decay mode allows the 
discrimination between 
two extreme scenarios of 
a pure \cp-even (as in the SM)
and a pure \cp-odd tensor structure at a level of 4.5--5.3$\sigma$ using about
10~\ifb\ of data (assuming the production rate is that of the SM, which is in 
conflict with the latest search limits from the Tevatron \cite{Tevlimits}).
A discriminating power of 2$\sigma$ was declared in the $\tau^+\tau^-$ decay 
mode at $M_H=120$~GeV and luminosity of 30~\ifb.

The situation is different in MSSM:
for $M_H \approx M_A \gtrsim 2 M_W$ the lightest MSSM
Higgs boson couples to gauge bosons with about SM strength, but its mass
is bounded to a region $M_h \lesssim 135$~GeV \cite{mhiggsAEC},
where the decay to
$WW^{(*)}$ or $ZZ^{(*)}$ is difficult to exploit.
On the other hand, the heavy MSSM Higgs bosons decouple from the 
gauge bosons. Consequently, since the usually quoted results for the 
$H \rightarrow ZZ/WW \rightarrow 4l$ channels assume a relatively 
heavy ($M_H \gtrsim 135$~GeV) SM-like Higgs, these results are not
applicable to the case of the MSSM. The above mentioned analysis of the weak
boson fusion with $H\rightarrow\tau^+\tau^-$ is applicable to the light 
\cp-even Higgs boson in MSSM but due to insignificant enhancements compared 
to the SM case no improvement can be expected. 

An alternative method which does not rely on the decay into a pair of gauge
bosons or on the production in weak boson fusion would therefore be of great
interest. Thanks to the $J_z=0$, C-even, P-even selection rule, the CEP Higss 
boson production in MSSM can yield a direct information about spin and \cp\
properties of the detected Higgs boson candidate. It is also expected, in 
particular in a situation where a new particle state has also been detected in 
one or more of the conventional Higgs search channels, that already a small 
yield of CEP events will be sufficient for extracting relevant information on 
the spin and $\cp$-properties of the new state \cite{HKRTW}.


\subsection*{Acknowledgments}
MGR thanks the IPPP at the University of Durham for hospitality. 
The work by MGR was supported by the Federal Program of the Russian State
RSGSS-3628.2008.2. This work is also supported in part by the network 
PITN-GA-2010-264564 (LHCPhenoNet). 
The work of MT was supported by the 
project AV0-Z10100502 of the Academy of Sciences of the Czech republic and
project LC527 of the Ministry of Education of the Czech republic.
The work of S.H. was supported in part by CICYT (grant FPA 2010--22163-C02-01) 
and by the Spanish MICINN's Consolider-Ingenio 2010 Program under grant 
MultiDark CSD2009-00064.

\end{document}